\documentclass[12pt]{article}
\usepackage[T1]{fontenc}
\usepackage{amssymb,amsthm,amsmath,mathrsfs,bm}

\begin{document}

\title{{ The Even and Odd Supersymmetric Hunter -  Saxton and Liouville Equations}}

\maketitle

   \hspace{2cm} $ Q.~P.~Liu^1, Ziemowit ~Popowicz^2, Kai ~Tian^3 $
\vspace{1.0cm}

$^{1}$ Department of Mathematics, China University of Mining and Technology,

Beijing 100083, P. R. China

 $^{2}$  Institute of Theoretical Physics, University of Wroc{\l}aw, pl. M. Borna 9,

50-205, Wroc{\l}aw, Poland

 $^{3}$  LSEC, ICMSEC, Academy of Mathematics and Systems Science,

Chinese Academy of Sciences, Beijing 100190, P. R. China

\hspace{4cm}

%
\begin{abstract}
It is  shown that two different  supersymmetric extensions of the Harry Dym
equation  lead to two different negative hierarchies of the supersymmetric integrable equations.
While the first one yields the known even supersymmetric Hunter - Saxton equation, the second one is a new odd  supersymmetric Hunter - Saxton equation.
It is further proved that these two  supersymmetric extensions of the Hunter - Saxton equation are reciprocally transformed to two different supersymmetric extensions of the Liouville equation.
\end{abstract}

\section{Introduction}
The list of unusual behavior of the supersymmetric  integrable systems is rather
long. It appears that during the supersymmetrization of the classical systems, some
typical supersymmetric effects, compared with the classical
theory, occur. The non-uniqueness of the roots of the supersymmetric Lax
operator \cite{pop1}, the lack of the bosonic reduction
to the classical equations \cite{krzywy} and the occurrence of the non-local
conservation laws \cite{nonl1,nonl2} have been discovered in the last century.
Recently it appeared that both  even and odd Hamiltonian operators could be
used to the  supersymmetrization of the classical integrable systems  \cite{susHD,recip,Liu,pop2}.
These effects rely strongly on the descriptions of the generalized classical systems
which we would like to supersymmetrize.

In this paper we prolong this list, namely we show that the Hunter -
Saxton (HS) equation, similarly as the Harry Dym (HD) equation, could be supersymmetrized in two different manners.
It is known that the classical HD equation is supersymmetrized in two ways, either by even or by odd supersymmetric
Hamiltonian operators \cite{susHD, recip}. It should be remarked that in addition to these two cases, 
a super HD equation was deduced from the fermionic extension of energy-dependent Schr\"{o}dinger operator \cite{Allan}. 
%

The even supersymmetric HD equation is a multi - Hamiltonian system and its
negative hierarchy contains the even supersymmetric generalization of the HS
equation. The second Hamiltonian structure for this extension
is generated by the supersymmetric  centerless Virasoro algebra.
In the second approach, where we use the odd Hamiltonian operators, an odd
supersymmetric HD equation is obtained from the Lax representation. However this equation admits bi - Hamiltonian
formulation only and does not possess  the `first' Hamiltonian operator.

As we shall show that, from the knowledge of the second and third odd supersymmetric
Hamiltonian structures of the odd supersymmetric HD equation,  it is
possible to construct new negative hierarchy  of equations. The flows of this hierarchy
contain a  new supersymmetric extension of the HS equation.

It is well known that the classical HS equation under reciprocal transformation
reduces to the Liouville equation \cite{hhdai}. We show  that it is possible to apply  the supersymmetric analogue of reciprocal  transformation described in \cite{recip} to two different supersymmetric extensions of the
HS equation, and as a result two supersymmetric extensions of the Liouville equation are obtained.

The paper is organized as follows. In the first section we recapitulate the known
facts on the bi - Hamiltonian formulation of the
classical HS equation and explain its connections with the
negative hierarchy of the HD equation.
In the next section the even hierarchy of the supersymmetric HD
equation as well as its negative hierarchy, which
contains the even supersymmetric extension of the HS equation, are
presented. Also this section describes the reciprocal link between the even supersymmetric
HS equation and a new supersymmetric Liouville equation.
The third section contains our main result where  the negative hierarchy of the odd supersymmetric HD equation is introduced. This hierarchy  takes an odd
supersymmetric HS equation as one of its flows.  Similar to the previous section
we describe a reciprocal link between this odd supersymmetric HS equation and a new supersymmetric Liouville equation.
In the appendix a simple proof is presented for the fact that the third Hamiltonian operators of the
even as well odd supersymmetric HD equations do
satisfy the Jacobi identity.

\section{Hunter - Saxton equation}
The Hunter - Saxton equation \cite{hs}
\begin{equation}\label{hs1}
u_{xxt}  =   2u_xu_{xx} + uu_{xxx}
\end{equation}
could be constructed in two different approaches. In the first approach this
equation is considered as the special limit of the
Camassa - Holm equation \cite{hhdai} while in the second  one it is regarded  as a member of the
negative hierarchy  of  the Harry Dym equation.

Let us briefly recapitulate these scenarios.

Both the HS equation and the Camassa - Holm equation could be embedded into the general model
\begin{equation}\label{ghs}
\lambda u_t - u_{xxt} = \frac{1}{2}\Big ( -3\lambda u^2 + 2uu_{xx} + u_x^2 \Big)_x,
\end{equation}
where $\lambda$ is an arbitrary parameter. For $\lambda =1$ the equation
\eqref{ghs} is just the Camassa - Holm equation \cite{comas, comas1}while
for $\lambda = 0$ it becomes the HS equation \eqref{hs1}.

The general equation \eqref{ghs} is a bi-Hamiltonian system
\begin{equation*}
m_t = K_1\frac{\delta H_1}{\delta m} =K_2\frac{\delta H_2}{\delta m},
\end{equation*}
where $m=\lambda u  - u_{xx} $ and
\begin{align*}
&H_1=\frac{1}{2}\int {\tt d}x\; um, && H_2=-\frac{1}{2}\int {\tt d}x\;(\lambda u^3 + uu_x^2),\\
&K_1 =- (\partial_x m  + m \partial_x ), && K_2=(\lambda -\partial_{x}^2)\partial_x.
\end{align*}

To put the HS equation into the negative HD hierarchy, we first consider the bi - Hamiltonian  structure of
the HD equation
\begin{equation}  \label{harry}
 w_t=P_1\frac{\delta H_{-1}}{\delta w} = P_2\frac{\delta H_{-2}}{\delta w} =
(w^{-\frac{1}{2}})_{xxx}
\end{equation}
where
\begin{align}
& P_1=\partial_x^3, && P_2= \partial_x w + w\partial_x=2w^{\frac{1}{2}} \partial_x w^{\frac{1}{2}}, \nonumber \\
& H_{-1}= 2\int {\tt d}x\; w^{\frac{1}{2}},  && H_{-2} = \frac{1}{8}\int {\tt d}x\;w^{-\frac{5}{2}} w_x^2. \label{ccas}
\end{align}

As we see in both approaches we deal with the same bi - Hamiltonian
structure.
In fact the HD equation possesses the multi -  Hamiltonian structure which is
constructed out of the recursion operator $R=P_2P_1^{-1}$ and $P_2 $
and reads as $P_{n+2}=R^nP_2, n=1,2,\dots$. For example the third Hamiltonian structure is
\begin{equation*}
 w_t=P_3 \frac{\delta H_{-3}}{\delta w}
\end{equation*}
where
\begin{align*}
&P_3=P_2 P_1^{-1} P_2 = (\partial_x w + w\partial_x )\partial_x^{-3} (\partial_x w +
w \partial_x), \\
&H_{-3}=-\frac{1}{2} \int {\tt d}x \; \big(16w_{xx}^2w^{-\frac{7}{2}} -
35w_x^{4}w^{-\frac{11}{2}}\big).
\end{align*}

Using the Hamiltonian operators $P_1$ and $P_2$, it is possible to construct the
negative as well as positive hierarchies of flows.
For the negative hierarchy, the first three flows are
\begin{eqnarray}
 w_{t_1}&=&P_1\frac{\delta H_{0}}{\delta w} = P_2\frac{\delta H_{-1}}{\delta w} =
P_3\frac{\delta H_{-2}}{\delta w}=0, \nonumber \\
w_{t_2}&=&P_1\frac{\delta H_{1}}{\delta w} = P_2\frac{\delta H_{0}}{\delta w} =
w_x  \label{drugi}, \\
w_{t_3}&=&P_1\frac{\delta H_{2}}{\delta w} = P_2\frac{\delta H_{1}}{\delta w}
=P_3\frac{\delta H_0}{\delta w}=
-w_x(\partial_x^{-2}w) - 2w(\partial_x^{-1}w),  \label{hs2}
\end{eqnarray}
where
\begin{equation} \label{ccas1}
 H_0= \int {\tt d}x \; w\quad, H_1 =\frac{1}{2} \int {\tt d}x \; (\partial_x^{-1}w)^2, \quad
H_2= \frac{1}{2}\int {\tt d}x \; (\partial_x^{-2} w)(\partial_x^{-1}w)^2.
\end{equation}

The equation \eqref{hs2} is the HS equation \eqref{hs1} after
identifying $t_3 = t,w=u_{xx}$.
Notice that $w_{t_2}$ is not a tri-Hamiltonian because
\begin{equation*}
 P_3^{-1}w_x = \frac{1}{4}
w^{-\frac{1}{2}}\partial_x^{-1}w^{-\frac{1}{2}}\partial_x^{3}
w^{-\frac{1}{2}}\partial_x^{-1}w^{-\frac{1}{2}} w_x =0.
\end{equation*}

The HS equation \eqref{hs1} is connected with the Liouville equation by a reciprocal transformation. Indeed
let us rewrite this equation in the conservative form as
\begin{equation*}
 v_t = (v\partial^{-2} v^2)_{x}
\end{equation*}
where $v^2 = u$.
This form implies the reciprocal transformation as
\begin{equation*}
 {\tt d} y= v {\tt d}x + (v\partial^{-2}v^2) {\tt d}t , \qquad {\tt d}\tau = {\tt d} t
\end{equation*}
and therefore
\begin{equation*}
 \partial_{x} = v \partial_{y}, \qquad \partial_{t} = \partial_{\tau} + (v \partial^{-2}_{x} v^2) \partial_{y}.
\end{equation*}
It is straightforward to show that  the resulted equation reads as
\begin{equation*}
 (\log v)_{y\tau} = v,
\end{equation*}
which is the celebrated the Liouville equation.

\section{Even Supersymmetric Hunter - Saxton equation}

The even supersymmetric HD equation has been
constructed by the supersymmetrization of
two Hamiltonian operators $P_1$ and $P_2$ in \cite{susHD}. The supersymmetric partner of the
second Hamiltonian operator $P_2$ corresponds in fact to
the supersymmetric centerless Virasoro algebra. The supersymmetric
Hamiltonian structures are given by
\begin{eqnarray*}
 \hat K_1 &=& {\cal D} \partial_x^2, \\
\hat K_2 & =& \frac{1}{2} [ W \partial_x + 2\partial_x W + ({\cal D}W){\cal D} ],
\end{eqnarray*}
where  ${\cal D}=\partial_\theta + \theta \partial_x$ and $W=\chi(x,t) +\theta u(x,t)$ is
a fermionic super field.

In these variables the even supersymmetric HD equation is
\begin{eqnarray}\label{susyHD}
W_t &=& \hat K_1\frac{\delta H_{-1}}{\delta w} =\hat K_2\frac{\delta
H_{-2}}{\delta W} \nonumber\\
&=&
\frac{1}{4}\partial_x^{2}\Big[ - 4 W_x({\cal D}W)^{-\frac{3}{2}} + 3W({\cal
D}W_x)({\cal D}W)^{-\frac{5}{2}}\Big],
\end{eqnarray}
where
\begin{eqnarray*}
 H_{-1}&=& \int {\tt d}x {\tt d}\theta\; W({\cal D}W)^{-\frac{1}{2}} =\frac{1}{2}\int {\tt d}x
\Big(2u^{-\frac{1}{2}}  -\chi_x\chi u^{-\frac{3}{2}}\Big), \\
H_{-2}&=&\frac{1}{16}\int {\tt d}x {\tt d}\theta\;\Big( W_x({\cal D}W_x)({\cal
D}W)^{-\frac{5}{2}} - 15WW_xW_{xx}({\cal D}W)^{-\frac{7}{2}}\Big)\\
&=& \frac{1}{16} \int {\tt d}x \;\Big( u_x^2u^{-\frac{5}{2}} +
16\chi_{xx}\chi_xu^{-\frac{5}{2}} -15\chi_{xx}\chi u_xu^{-\frac{7}{2}}+
15\chi_x\chi u_{xx}u^{-\frac{7}{2}} \Big).
\end{eqnarray*}
In components the equation \eqref{susyHD} reads
\begin{eqnarray*}
 \chi_t &=& -\frac{1}{2}\Big ( (\chi u^{-\frac{3}{2}})_x + \chi_{x} u^{-\frac{3}{2}} \Big
)_{xx}, \\
u_t &=& \frac{1}{4}\Big (2u^{-\frac{1}{2}} + 3\chi_x\chi u^{-\frac{5}{2}} \Big )_{xxx}.
\end{eqnarray*}

In fact this supersymmetric generalization constitutes the  tri-Hamiltonian
system also, i.e.
\begin{equation}
 W_t  = \hat K_3 \frac{\delta H_{-3}}{\delta W} =\hat K_2 \hat K_1^{-1} \hat
K_2 \frac{\delta H_{-3}}{\partial W },
\end{equation}
where
\begin{eqnarray}\nonumber
H_{-3} &=& -\frac{1}{384} \int {\tt d}x {\tt d}\theta \; W \Big ( 420 W_{xxx}W_{xx}({\cal
D}W)^{-\frac{9}{2}} +2394 W_{xx}W_{x} ({\cal D} W_{xx})({\cal D}W)^{-\frac{11}{2}} \\ \nonumber
&& - 875({\cal D}W_{xxx}) ({\cal D}W_x)({\cal D}W)^{-\frac{9}{2}}
+ 7({\cal D}W_{xx})^{2}({\cal D}W)^{-\frac{9}{2}} +
348({\cal D}W_{4x})({\cal D}W)^{-\frac{7}{2}} \Big ).
\end{eqnarray}
The proof that $\hat K_3$ satisfies the
Jacobi identity is postponed to the appendix.

The negative hierarchy of the supersymmetric HD equation \cite{susHD} is
defined as in the  classical case. It reads as
\begin{eqnarray}
W_{t_1} &=&\hat K_1\frac{\delta H_{0}}{\delta W} = \hat K_2\frac{\delta
H_{-1}}{\delta W} = \hat K_3\frac{\delta H_{-2}}{\delta W}=0, \nonumber\\
W_{t_2}&=&\hat K_1\frac{\delta H_{1}}{\delta W} = \hat K_2\frac{\delta H_{0}}{\delta W} =
W_x  \label{drugi}, \\
W_{t_3}&=&\hat K_1\frac{\delta H_{2}}{\delta W} = \hat K_2\frac{\delta H_{1}}{\delta W} =
\hat K_3\frac{\delta H_0}{\delta W} \nonumber \\
&=& -\frac{3}{2} W ({\cal D}^{-1}W)  -W_{x} ({\cal D}^{-3} W)  -\frac{1}{2} ({\cal
D}W) (\partial_x^{-1} W), \label{shs1}
\end{eqnarray}
where
\begin{align*} \label{ccas1}
& H_0= \int {\tt d}x {\tt d}\theta  \; W,\qquad H_1 =-\frac{1}{4} \int {\tt d}x {\tt d}\theta \; ({\cal
D}^{-3}W)W, \\
& H_2= \frac{1}{2}\int {\tt d}x{\tt d}\theta  \; ({\cal D}^{-1}W)(\partial^{-1} W)({\cal
D}^{-3}W).
\end{align*}

We remark here that the equation \eqref{shs1} is the even supersymmetric extension of the HS equation which has been considered first time in
\cite{susHD} and later rediscovered in \cite{lenel}.

This supersymmetric  equation could be be rewritten in the new bosonic super field
$U=-({\cal D}W)/2$ as
\begin{equation}\label{nshs1}
U_{t_3} = 2U_x(\partial_x^{-2} U) + 4U(\partial_x^{-1} U)  - ({\cal D}U) ({\cal D}^{-3}U),
\end{equation}
which has a conservation law
\begin{equation}\label{conshs1}
\Big(U^{\frac{1}{4}}\Big)_{t_3} = {\cal D} \Big ( U^{\frac{1}{4}} ({\cal D}^{-3}U)
+ 2({\cal D}U^{\frac{1}{4}}) (\partial_x^{-2} U) \Big ).
\end{equation}
In that way we prove that
\begin{equation*}
H= \int {\tt d}x {\tt d}\theta \; U^{\frac{1}{4}} = \frac{1}{4} \int {\tt d}x \; \xi q^{-\frac{3}{4}}
\end{equation*}
is a fermionic conserved quantity with the non  classical analog, where $U=q+\theta \xi$.

To construct a reciprocal transformation for the
equation \eqref{nshs1}, we turn to the Proposition 1 established in \cite{recip}. Thus, in addition to  the conservation law \eqref{conshs1} a potential
is required. In this case,
we have
\begin{equation*}
\mathcal{D}\Big(2U^{\frac{1}{2}} (\partial_x^{-2} U)\Big)=2U^{\frac{1}{4}}\Big ( U^{\frac{1}{4}} ({\cal D}^{-3}U)
+ 2({\cal D}U^{\frac{1}{4}}) (\partial_x^{-2} U) \Big ).
\end{equation*}
Hence, a reciprocal transformation is formulated as
\begin{align}
& \mathcal{D}=U^{\frac{1}{4}}\mathbb{D}, \label{rt11}\\
& \frac{\partial }{\partial t_3}= \frac{\partial}{\partial \tau} + 2U^{\frac{1}{2}} (\partial_x^{-2} U) \frac{\partial}{\partial y} +\Big ( U^{\frac{1}{4}} ({\cal D}^{-3}U)
+ 2({\cal D}U^{\frac{1}{4}}) (\partial_x^{-2} U) \Big )\mathbb{D}. \label{rt12}
\end{align}

Applying the transformation \eqref{rt11}-\eqref{rt12} we obtain a new supersymmetric generalization of the
Liouville equation
\begin{equation}
(\log U)_{y\tau} =4U^{\frac{1}{2}} - U^{-\frac{5}{4}}(\mathbb{D}U)(\mathbb{D}^{-1}U^{\frac{3}{4}}).
\end{equation}
To see the connection with Liouville equation, we rewrite above equation in terms of components. It yields
\begin{eqnarray}
(\log q)_{y\tau} & = & 4q^{1 \over 2} - \frac{3}{4}\eta q^{-\frac{1}{4}}(\partial_x^{-1}\eta q^{3 \over 4} ),\\
\eta_\tau & = & 3\eta q^{1 \over 2} - \frac{3}{4}q_xq^{-\frac{5}{4}} (\partial_x^{-1}\eta q^{3 \over 4} ),
\end{eqnarray}
where   $\eta=q^{-1}\xi$.
When $\eta =0$, this system reduces to the Liouville equation.

\section{Odd  Supersymmetric Hunter - Saxton Equation}

The other supersymmetric HD equation was worked out from  a Lax operator and the associated Lax representation \cite{susHD}. Indeed,
the supersymmetric  Lax operator
\begin{equation}
L = ({\cal D}W)^{-1} {\cal D}^4 - \frac{1}{2}W_x({\cal D}W)^{2} {\cal D}^3
\end{equation}
leads to the following generalization of the Harry Dym equation
\[
L_t = \Big [ L^{\frac{3}{2}}_{\geq 3}, L \Big ],
\]
or
\begin{eqnarray*}
 W_t &=& \frac{1}{16} \Big [ 8{\cal D}^5({\cal D}W)^{-\frac{1}{2}}    - 3{\cal
D}\big(W_{xx}W_x({\cal D}W)^{-\frac{5}{2}}\big) \\
&& + \frac{3}{4}({\cal D}W_x)^2W_x({\cal D}W)^{-\frac{7}{2}} -
\frac{3}{4} {\cal D}^{-1}\big(({\cal D}W_x)^3({\cal D}W)^{-\frac{7}{2}}\big) \Big ],
\end{eqnarray*}
where $W$ is a fermionic super field.
For our next purposes let us rewrite this equation by means of  $V=({\cal
D}W)^\frac{1}{2}$
as
\begin{equation}\label{oshd}
 V_t=-\frac{1}{8}\Big [ \partial_x^3 (V^{-2}) - 3{\cal D} \partial_x\big( ({\cal
D}V)V_xV^{-4}\big) \Big ].
\end{equation}
In the components   the last equation becomes
\begin{eqnarray*}
 u_t &=& -\frac{1}{8}\partial_x \Big(\partial_x^2 (u^{-2}) - 3 u_x^2 u^{-4} - 3 \chi_x\chi u^{-4}\Big), \\
\chi_t &=& \frac{1}{8}\partial_x^{2}  \Big( 2\partial (u^{-3} \chi) - 3u_xu^{-4}\chi  \Big),
\end{eqnarray*}
where $V=u(x,t)+\theta\chi(x,t)$.

The classical HD equation \eqref{harry} could be obtained in the bosonic
sector where $\chi =0$ and
$u=w^{1 \over 2}$. Thus both even and odd supersymmetric extensions contain the classical HD equation but
their fermionic extensions are different.

The bi-Hamiltonian structure for \eqref{oshd} was constructed very recently in \cite{recip},  it reads
\begin{equation}
 V_{t}=\hat{P_2} \frac{\delta \hat H_{-2}}{\delta V} = \hat{P_3}\frac{\delta \hat H_{-3}}{\delta V},
\end{equation}
where
\begin{align}
\hat H_{-2} = & -\frac{1}{8} \int  {\tt d}x {\tt d}\theta \; ({\cal D}V)V_xV^{-3} =
-\frac{1}{8} \int {\tt d}x \; (u_x^2u^{-3} + \chi_x\chi u^{-3} ) , \label{h2} \\
\hat H_{-3} = & - \int {\tt d}x{\tt d}\theta\;({\cal D}V)\big ( 4V_{xxx}V^{-5} - 30V_{xx}V_{x}V^{-6} +
35V_x^{3}V^{-7} \big )  \nonumber \\
= & \int {\tt d}x \; \Big (4u_{xx}^2u^{-5} - 15u_x^4u^{-7} + 4\chi_{xx}\chi_x u^{-5}-\chi_x\chi (45u_x^2u^{-7} - 20u_{xx}u^{-6})\Big ),\\
\hat{P_2} =& {\cal D}^3,  \\
\hat{P_3} =& V^{\frac{1}{2}}{\cal D} V^{-\frac{1}{2}}{\cal D}V{\cal
D}^{-5}V{\cal D}V^{-\frac{1}{2}}
{\cal D}V^{\frac{1}{2}}  \label{Pe3} \nonumber \\
= &  \frac{1}{4} \Big [ ({\cal D} V) \partial_x^{-2} (({\cal D}V){\cal D} -  V
\partial_x - \partial_x V) +
2\partial_x V\partial_x^{-3}( 2 V \partial_x {\cal D} - 2({\cal D}V_x) + 3\partial_x
({\cal D}V)) \Big ].
\end{align}
The above Hamiltonian operators in the components are given by
\begin{equation}
\hat{P_2}=\left( \begin{array}{cc}
             \partial_x & 0 \\
	0 & \partial_x^{2}
            \end{array}\right ),
\end{equation}
\begin{equation}
\hat{P_3}  =
\frac{1}{4} \left(
\begin{array}{cc}
  \partial_x u \partial_x^{-3} u \partial_x +\chi \partial_x^{-2} \chi  &
\begin{array}{c} -\chi \partial_x^{-1}(\partial_x^{-1} u \partial_x + u) + \\
2\partial_x u \partial_x^{-2} (-2\partial_x^{-1}\chi_x + 3\chi)
\end{array}
\\[12pt]
 \begin{array}{c} -(u + \partial_x u \partial_x^{-1}) \partial_x^{-1} \chi + \\
2(2\chi_x\partial_x^{-1} + 3\chi)\partial_x^{-2}u\partial_x
\end{array}   &
\begin{array}{c} u^2 + u\partial_x^{-1}u\partial_x +\partial_x u\partial_x^{-1}u +
\partial_x u \partial_x^{-2}u\partial_x
\\ + (\chi + 2\partial_x \chi \partial_x^{-1} ) \partial_x^{-1} (2\partial_x^{-1} \chi
\partial_x + \chi)
\end{array}
 \end{array} \right).
\end{equation}

In passing, we notice that the Poisson brackets induced by  $\hat{P_2}$ and $\hat{P_3}$ operators are given by
\begin{equation}
 \{ V(x,\theta),V(y,\theta^{'}) \} = \hat{P_i} ~ \delta(x-y) ~ (\theta - \theta^{'}) ,
~~~~ i=2,3.
\end{equation}
As $V$ is a bosonic super field and $\hat{P_i}$ is an superfermionic operator hence
we are dealing with the so called odd Poisson brackets.

The Jacobi identity for the operator $P_3$ is proved in \cite{recip} and in the appendix we present a quite different verification   of this property.

In contrast to the classical case we do not know the first Hamiltonian operator
for  the odd supersymmetric HD equation \eqref{oshd}.
However we can  construct the negative hierarchy using our bi - Hamiltonian
structure only.

To this end, let us first find the supersymmetric analog  of the classical
Casimir functions for the equation \eqref{oshd}.
In the classical case  $H_{-1} = \int dx ~ u $ and $H_0=\int dx ~ u^2$  defined
by \eqref{ccas}, \eqref{ccas1} are the Casimir
functions for the Hamiltonians operators $P_2$ and $P_1$  respectively while $H_0$ is
a Casimir function for $P_3$.

In the supersymmetric case we find out  that
\begin{equation}
\hat H_{-1} = \int {\tt d}x {\tt d}\theta \; V^{-1} ({\cal D}^{-1} V^2) = \int {\tt d}x \;
\big(u - 2\chi u^{-2}\partial_x^{-1} \chi u\big)
\end{equation}
 is conserved quantity for the equation \eqref{oshd} and is a
Casimir function for $\hat P_2$.
On the other side,  $\hat H_{-2}$ defined by \eqref{h2} is the Casimir function for
$\hat P_3$.

Interestingly we have an additional  new
Casimir function for the operator $\hat P_2$
\begin{equation}
 \hat G_{0}=\int dx d\theta ~ V =\int dx ~ \chi
\end{equation}
whose parity is different from that of $\hat H_{-1}$ (or $\hat H_{-2}$). If we use $\hat G_0$ to
the construction of the negative hierarchy it appears that this hierarchy will
be of a different parity than the odd supersymmetric HD equation \eqref{oshd} and from the physical point of view it may not be
 relevant.  Indeed if we consider the following equation
\begin{equation}
 V_{\varsigma}=P_3\frac{\delta \hat G_0}{\delta V} = \frac{1}{4} \big [ 2\partial_x V
\partial_x^{-2} {\cal D}V -
({\cal D}V)\partial_x^{-1}V \big ],
\end{equation}
which in the components reads
\begin{eqnarray*}
 u_{\varsigma} &=& \frac{1}{4} \big [ -\chi \partial_x^{-1} u + 2\partial_x u
\partial_x^{-2} \chi \big ], \\
\chi_{\varsigma} &=& \frac{1}{4} \big [ u^2 + \chi \partial_x^{-1} \chi + \partial_x u
\partial_x^{-1} u + 2\partial_x \chi\partial_x^{-2} \chi \big ],
\end{eqnarray*}
then we have to assume that $\varsigma$ is a fermionic time.

From $\hat H_{-1}$ and $\hat H_{-2}$ we obtain the first member of the
negative hierarchy of the odd supersymmetric HD equation as
\begin{equation}
 V_{t_1}= \hat P_2\frac{\delta \hat H_{-1}}{\delta V} = \hat P_3\frac{\delta \hat H_{-2}}{\delta V} =0.
\end{equation}

In the classical case the second member of the negative hierarchy of HD
equation is not tri - Hamiltonian but bi - Hamiltonian
system generated by the operators $P_1,P_2$ and Hamiltonians $H_1$ and $H_0$ only.
In order to construct the second member of the negative hierarchy in the
supersymmetric case let us
 build  the superfermionic partner of  the classical $H_0= \int dx ~w=\int dx ~
u^2 $ as in the equation \eqref{ccas1}.
The result is
\begin{equation}
 \hat H_0=\int {\tt d}x {\tt d}\theta \; V ({\cal D}^{-1} V) = \int {\tt d}x \; \big(u^2 + \chi (\partial^{-1}
\chi)\big).
\end{equation}
It is a conserved quantity for the odd supersymmetric HD equation \eqref{oshd} but not a
Casimir function $ \hat P_2$ or $\hat P_3$. This Hamiltonian generates the following translation flow
\begin{equation}
 V_{t_2}= \hat P_2\frac{\delta \hat H_0}{\delta V} = V_x
\end{equation}
which gives us in the bosonic limit the second negative flow of the classical
HD hierarchy \eqref{drugi}.

Now the  conserved quantity $H_0$ generates the supersymmetric analog of the
HS equation which  is  a bi - Hamiltonian
system also
\begin{eqnarray}
 V_{t_3} &=& \hat P_3\frac{\delta \hat H_0}{\delta V} =\hat P_2\frac{\delta \hat H_1}{\delta V} \nonumber\\
 &=&
\partial_x\Big( V  {\cal D}^{-3}[ V({\cal D}^{-1}V)]\Big) - \frac{1}{2}({\cal
D}V)\partial_x^{-1}[V({\cal D}^{-1}V)],\label{susHS}
\end{eqnarray}
where
\begin{align*}
\hat H_1 =& \int {\tt d}x {\tt d}\theta \;\big [2({\cal D}^{-1}V) V {\cal D}^{-3}  -
({\cal D}V)({\cal D}^{-3}V)\partial^{-1}\big ]V{\cal D}^{-1}V \\
= &\int {\tt d}x ~\big [2u^2\partial^{-2} + 2\chi(\partial^{-1}\chi)\partial^{-2} -
\chi(\partial^{-2}\chi)\partial^{-1} \big ]
(u^2+\chi\partial^{-1}\chi) - \\ \nonumber
& \hspace{2.5cm} 2(\partial^{-1}\chi)u\partial^{-1}u\partial^{-1}\chi +
(\partial^{-2}\chi)u^2\partial^{-1}\chi.
\end{align*}
In components we have
\begin{align*}
u_{t_3} =& \partial_x u \partial_x^{-2}(u^2 + \chi\partial_x^{-1}\chi) -
\frac{1}{2}\chi\partial_x^{-1}u\partial_x^{-1}\chi, \\
\chi_{t_3} =&\chi_x\partial_x^{-2}(u^2+\chi\partial_x^{-1}\chi)
+\frac{3}{2}\chi\partial_x^{-1}(u^2+\chi\partial_x^{-1}\chi)
+\frac{1}{2}u_x\partial_x^{-1}u\partial_x^{-1}\chi+u^2\partial_x^{-1}\chi.
\end{align*}
If we take $\chi=0$ and $w=u^2$,  then $w_{t_3}$ is nothing but the
HS equation.

Now we turn to the construction of  possible reciprocal transformation for the supersymmetric HS equation \eqref{susHS}.
It is observed that the equation \eqref{susHS} has the following  conservation law
\begin{equation}
 \Big ( V^{\frac{1}{2}}\Big )_{t_3} = \frac{1}{2}\frac{\partial }{\partial x}\Big ( V^{\frac{1}{2}}
{\cal D}^{-3} [ V({\cal D}^{-1} V )] \Big ) + \frac{1}{4} {\cal D} \Big( V^{-\frac{1}{2}} ({\cal D}V)
{\cal D}^{-3} [ V ( {\cal D}^{-1} V ) ] \Big).
\end{equation}
Furthemore, a potential can be introduced for the product of the conserved density and flux in
the above conservation law
\begin{align*}
\mathcal{D}\Big(V\mathcal{D}^{-3}[V(\mathcal{D}^{-1}V)]\Big)
=V^{1\over2}\mathcal{D}\Big(V^{1\over2}\mathcal{D}^{-3}[V(\mathcal{D}^{-1}V)]\Big)
+\frac{1}{2}(\mathcal{D}V)\mathcal{D}^{-3}[V(\mathcal{D}^{-1}V)].
\end{align*}
Hence, we have a reciprocal transformation given by
\begin{align}
\mathcal{D}=&V^{1\over2}\mathbb{D},\\
\frac{\partial}{\partial t_3}=&\frac{\partial}{\partial \tau}+V\mathcal{D}^{-3}[V(\mathcal{D}^{-1}V)]\frac{\partial}{\partial y}\nonumber\\
&+\frac{1}{2}\Big(V^{-\frac{1}{2}}(\mathcal{D}V)\mathcal{D}^{-3}[V(\mathcal{D}^{-1}V)]
+V^{1\over2}\partial_x^{-1}[V(\mathcal{D}^{-1}V)]\Big)\mathbb{D},
\end{align}
which could be applied  to the supersymmetric HS equation \eqref{susHS}. A direct calculation gives us
\begin{equation}
V_\tau=V\mathbb{D}^{-1}[V^{1\over2}(\mathbb{D}^{-1}V^{1\over2})].
\end{equation}
By introducing $V=({\mathbb{D}}\Phi)^{2}$, the last equation is rewritten in a neat form
\begin{equation}
 2\frac{\partial }{\partial \tau} \mathbb{D} \log (\mathbb{D}\Phi) = (\mathbb{D}\Phi) \Phi
\end{equation}
which is a new supersymmetric generalization of the Liouville equation. To see it, we just rewrite this equation in components $\Phi = \xi + \varrho u$ as
\begin{eqnarray*}
 2(\log u )_{y\tau}&=& u^2 + \xi_y\xi, \\
2(u^{-1}\xi_y)_{\tau} &=&  u\xi,
\end{eqnarray*}
when $\xi =0$ this system goes back to the Liouville equation.

\section*{Acknowledgement}

 This
work is supported by National Natural Science Foundation of China
with grant numbers 10731080 and 10971222.

\section*{Appendix}

In order to prove that the operator $\hat K_3= \hat K_2 \hat K_1^{-1} \hat K_2$
satisfies the Jacobi identity we use the
idea of the decompression of the Hamiltonian operators  described in \cite{pop3}.
To this end  let us embed the $\hat K_3$ operator into the two dimensional matrix
as
\begin{equation*}
 J=\left(
\begin{array}{cc}
  2V_x + 3V \partial + ({\cal D}V) {\cal D}   &  \hat K_2
\\  \hat K_2 & \hat K_1 + 2V_x + 3V \partial + ({\cal D}V) {\cal D}
\end{array} \right )
\end{equation*}
where $V$ is a new superfermionic function.

This new $J$ operator does not contain the nonlocal terms hence it is  easy to
prove that the Jacobi identity holds for it indeed. Now using the Dirac reduction procedure one can easily
recognize that in the case $V=0$ the
 operator $J$ reduces to  $\hat K_3$, hence the last operator satisfies the
Jacobi identity also.

For the odd  operator $\hat P_3$, we reformulate it  as
\begin{equation*}
\hat P_3 = V^{\frac{1}{2}}{\cal D} V^{-\frac{1}{2}}{\cal D}V{\cal D}^{-5}V{\cal
D}V^{-\frac{1}{2}}
{\cal D}V^{\frac{1}{2}} = - Q {\cal D}^{-5}  Q^{\star},
\end{equation*}
where
\[
Q = V^{\frac{1}{2}}{\cal D} V^{-\frac{1}{2}}{\cal D}V.
\]

Similarly to the previous case let us embed $\hat P_3$ operator to the graded two
dimensional matrix
\begin{equation*}
 \hat J=\left(
\begin{array}{cc}
 0  &  -\frac{1}{2} Q \\[10pt]
   \frac{1}{2} Q^{\star} ~~  &  {\cal D}^{5} - 2 W_x - 3W\partial - ({\cal
D}W){\cal D}
\end{array} \right )
\end{equation*}
where $W$ is a new superfermionic function.

This matrix generates the following Poisson brackets for the fields $U_i $
\begin{equation*}
\{  U_i(x,\theta),U_j(y,\theta^{'}) \} = \hat J(i,j) \delta(x-y)(\theta -
\theta^{'}), ~~~ i,j=1,2
\end{equation*}
with identifications $U_1=V,U_2=W$.
Remember that the test functions are graded vectors it is easy to verify that $\hat J$ does satisfy the Jacobi identity. Applying the Dirac reduction procedure
with $W=0$, we find that the reduced operator is just $ \frac{1}{4} \hat P_3$, which  satisfies the
Jacobi identity too.

\end{document}